\documentclass{article}
\usepackage{graphicx} % Required for inserting images
\usepackage{amsmath}
\usepackage{amsthm}
\usepackage{amssymb}
\usepackage{authblk}
\usepackage{url}
\usepackage{comment}
\usepackage{mathtools}

\author[1]{Akaki Mamageishvili}
\author[2]{Christoph Schlegel}
\author[3]{Ko Sunghun}
\author[3]{Jinsuk Park}
\author[4]{Ali Taslimi}

\affil[1]{Offchain Labs}
\affil[2]{Flashbots}
\affil[3]{Matroos}
\affil[4]{Entropy Advisors}

\date{November, 2025}

\begin{document}
\title{TimeBoost: Do Ahead-of-Time Auctions Work?}
\maketitle

\begin{abstract}
    We study the performance of the TimeBoost auction, by comparing cumulative fixed time markout of fast lane trades over the TimeBoost interval to bids for the fast lane. Such comparison allows us to assess how well bids predict future extracted value from the time advantage. The correlation between winning bids and markouts is weak across bidders, suggesting that bids are a noisy predictor of extracted value. The correlation slightly improves when comparing paid bids (the second highest bid) instead of winning bids to markouts, which we attribute to the fact that the auction is more of a common value type. 
    In all settings, the relative order of the most frequent bidder performance remains the same, together with their absolute profits. Bids and markouts aggregated over long time intervals exhibit much higher correlation, indicating that bidders detect trends much better than identify when the high arbitrage value is exactly available. One possible explanation for this is the fact that the correlation between previous minute markouts and current minute bids is significant, suggesting that the previous minute markouts is used to predict the next minute value when bidding. 
\end{abstract}

\section{Introduction}
Many blockchains use auctions as part of their transaction ordering mechanism. In Ethereum, for example, {\it Proposer Builder Separation (PBS)} uses a builder auction where specialized parties, so called {\it block builders}, bid for the right to determine the content of the next proposed block.  In most rollups, transactions are sorted in decreasing priority fees per gas. This is commonly called {\it Priority Ordering} or {\it Priority Gas Auction}, referring to the fact that it is an auction for earlier execution. 
Besides significant differences --in Ethereum builders bid on the right to order the whole block whereas in rollups the auction is on the relative ordering of individual transactions by the sequencer, Ethereum has a public mempool, rollups do not-- these auction formats usually have one commonality: the auction is ''Just-in-Time" (JIT)  in the sense that bidders can bid in it just until immediately before the content of the block is determined. As an alternative, ''Ahead-of-Time" (AOT) auctions would have a sizable time gap between the auction and the determination of the order of transactions. 
In the context of Ethereum L1, these auctions have been discussed under the term ''slot auctions": in a slot auction, the builder only bids on the right to propose the next block without committing to any specific block content at the time of bidding and determines this content only later (usually 12 seconds later)\footnote{See \cite{slot_auction}.}. In rollups, TimeBoost implemented by Arbitrum~\footnote{https://docs.arbitrum.io/how-arbitrum-works/timeboost/gentle-introduction} is an instance of a transaction ordering policy that has an Ahead-of-Time component as part of its design.

To compare the relative performance of AOT and JIT auctions, it is important to assess how well bids match the value that bidders get from the right they acquire. In JIT auctions, this is just a question of competition. As long as enough bidders compete in the auction the paid bids should be close to the value that the winning bidders get. For AOT auctions, besides competition, there is also a problem of prediction. Bidders need to predict the future value that they will get. If they are able to predict it well (and there is sufficient competition), the auction will yield efficient outcomes. If they do not, the auction will generally be more inefficient than a JIT auction.

In this paper, we provide the first empirical analysis of an AOT auction in the context of blockchain transaction ordering, using Arbitrum's TimeBoost auction as a case study.
\paragraph{Related Work}
In Ethereum, the next block proposer sorts transactions in the block unilaterally. This allows the proposer to extract the value that comes from the power of ordering transactions. Such value is usually referred as Maximum Extractable Value, or MEV for short~\cite{MEV}.
Transaction ordering through auctions has been theoretically analyzed in~\cite{schlegel2024transaction}. An early version of TimeBoost (without an Ahead-of-Time auction) has been proposed in \cite{mamageishvili_et_al:LIPIcs.AFT.2023.23}.
Evaluating time advantage of arbitrage between centralized and decentralized exchanges (CEX-DEX arbitrage) under a First-Come-First-Serve (FCFS) ordering policy has been studied ~\cite{mev_capture}, comparing it to PGA auctions and pure FCFS. The latency advantage in a PBS setting has been studied in a series of papers~\cite{latency_advantage},~\cite{timing_games}. Possible downsides of AOT auctions have been identified in~\cite{pai2024centralization}, in case that there is a working secondary market at which the AOT acquired right can be resold (Just-in-Time).

Methodologically, we rely on the following sources: Markouts at centralized exchange mid prices have been used in many studies to estimate CEX-DEX arbitrage profits on Ethereum, e.g., in~\cite{cex_dex}. Theoretical predictions of CEX-DEX arbitrage profits have been made in~\cite{milionis2022automated}. Our theoretical background for auctions with common value component comes from the classical work of~\cite{auction_theory}.

\paragraph{TimeBoost}
TimeBoost is a relatively new transaction ordering policy deployed on Arbitrum rollup since April 17th, 2025. In this policy, a protocol sells a time advantage in the FCFS policy to one party in a permissionless manner, via a second-price auction. Bids are denominated in the Ethereum native asset, ETH. There is a reserve price of 0.001 ETH. If no bidder bids more than the reserve price, transaction ordering policy remains pure FCFS. If only one bidder bids more than the reserve price, it pays the reserve price. If at least two bidders bid more than the reserve price, then the winner is the highest bidder and it pays the second highest bid. The time advantage equals 200ms and it lasts 1 minute for the winner.
The auction for the next minute starts at second 0 of each minute and lasts until second 45. This introduces a 15 second time window for the bidders for predicting next minute's value. The reason for having a 15 second time window is to allow enough time for the auction settlement on-chain. 
The rule is enforced by a designated party, the sequencer, which executes regular transactions with a 200ms delay, while transactions coming from the TimeBoost winner are executed on their arrival, without delay. In effect, the sequencer merges these two FCFS queues into each other. The advantage, sometimes referred to as {\it fast lane}, the TimeBoost auction winner is given allows the party to react to external information faster than other users and e.g.~to take advantage of CEX-DEX arbitrage. TimeBoost also allows the party to react to large price changes on the chain faster than others.
Empirically, most of the extracted values and most activity in the fast lane can be attributed to the CEX-DEX arbitrage~\cite{essias2025express}. In particular, Wintermute exclusively extracts CEX-DEX arbitrage, while Selini sometimes uses fast lane for other types of arbitrage.

\paragraph{Empirical Approach}
For each winning bidder, we keep track of all transactions they submit through a fast lane at the on-chain exchanges. We calculate the hypothetical profit that the party would make if they would sell assets they bought on an external market a few seconds after, a profit referred to as a markout. The methodology likely leads to an overestimation, as average prices taken at CEXes may not reflect the real prices auction participants execute their trades at. However, we treat all players in this regard in the same way.
Such a heuristic approach to estimating arbitrage value is needed since there is no good way to identify what arbitrageurs are doing on external markets in parallel to their on-chain trades.
Markout serves as an unbiased measure of potential profit. For each minute, we calculate cumulative markout of the winner, which gives a markout timeseries. This allows to compare different bidders to each other. We compare the markouts to winning bids and paid bids (or a reserve price, in case there was no second bid). The bids of players are a proxy of the value they expect to extract in the next minute. 
\paragraph{Empirical Findings}
Bids and markouts are weakly correlated. Paid bids and markouts are slightly stronger correlated than winning bids and markouts, but the correlation is still weak. 
 The explanation for a (slightly) stronger correlation can likely be attributed to a strong common-value component in the auctioned good.

There are multiple explanations for why the correlation is low. First, the auction ends 15 seconds before the round even starts, increasing the uncertainty of the value in the next minute. Second, estimating the future value over 1 minute is difficult. The correlation between bids and arbitrage profits for longer time windows is higher, i.e. bids for 1 minute of TimeBoost, are better correlated with 5,10, etc. minutes of future profit than 1 mn inute of future profits.
 In particular, we find that the Pearson correlation between aggregate markouts and bids is increasing and even crossing the threshold of 0.8 for 30 minutes. This suggests that the bidders predict trends of high CEX-DEX arbitrage value, but can not predict with high accuracy the profit in the next minute.

  As we show, these results are consistent with the theory that arbitrage profits are proportional to return variance of the crypto-currencies traded. As variance is easier to predict for longer time intervals, predicting a longer average of profits seems easier to predict, whereas profits fluctuate a lot and are hard to predict in the short run.
  
   This finding is also observed in other literature. For example,~\cite{ANDERSEN2011220} compared the performances of forecasting models with varying sampling frequencies and showed that market microstructural noise causes a lower signal-to-noise ratio, resulting in a reduced correlation between the forecast and the actual value.~\cite{Hautsch2012} discusses forecasting 15-minute volatilities, and although the models are different from those used in~\cite{ANDERSEN2011220}, comparing $R^2$, one can notice that the performance of intraday forecasting is worse than that of daily ones.

Spearman correlation is stronger than the Pearson correlation between bids and minute markouts for transactions with positive return, suggesting that players are better at guessing the monotonicity of the future value than guessing the magnitude of the change, which intuitively sounds reasonable as guessing only the direction of value changing is easier. Spearman correlation is considerably lower than the Pearson correlation when all transactions are considered. This further motivates us to consider only transactions with positive markouts.   

Sorting the three most frequent bidders by their correlation coefficients in all time intervals, types of bids and type of correlation coefficient results into the same ordering in the month of August, which is the focus of our data analysis. However, the advantage of Wintermute against Selini can be attributed to a missing non CEX-DEX arbitrage calculation. The ordering changes when considering a longer time interval, June-August. 

%The most counterintuitive finding of (co-)relating bids (paid or winning) to markout profits is that in shorter time intervals, correlation is lower than the full time interval, even if the shorter time interval ends earlier. It goes against the intuition that bidders should be able to predict immediate future value better than the more distant future value. The result suggest that either on high frequency arbitrage value is too noisy or that there is more idiosyncratic  private value extracted by the winner in the beginning of the TimeBoost intervals. The first hypothesis seems more plausible in our opinion.

\section{Theoretical Predictions}
\subsection{Common Value Second Price Auctions}
The TimeBoost auction has an obvious common value component: the value that bidders can extract from arbitrage throughout the TimeBoost interval is common to all bidders.
 Besides this common value, bidders in the TimeBoost auction may also have a private idiosyncratic value component for winning the auction: bidders might have private inventory constraints, they might have accumulated positions that they want to unwind, if they are market makers on external exchanges, trades and inventories accumulated through their market making activity might determine the private value. They might also have heterogeneous trading cost on external markets that would make them value positions accumulated on Arbitrum chain exchanges differently.
 
 A suitable model for bidding under common, as well as private value is the classical \cite{auction_theory} framework. Suppose bidders receive private signals, $X_i$ for $i=1,\ldots,n$ that are i.i.d., and the value of winning the auction for bidder $i$ is a function of all signals $v_i(X_1,\ldots,X_n)$. For the second price auction format of the TimeBoost auction it predicts that a bidder $i$ that receives a signal $X_i=x$  bids
 $$b_i:=E[v_i(x,X_{-i})|i\text{ wins}]=E[v_i(x,X_{-i})|\max_{j\neq i}X_j\leq x]<E[v_i(x,X_{-i})].$$
 Note that in contrast to a pure private value auction bidders do not bid their estimated value based on their signal (which would be the expression on the rhs of the inequality), but condition their bid value on winning which yields a strictly smaller bid. 

 The theory would make the following prediction:
 If the paid bid is a better predictor of the value than the winning bid, then this is consistent with the common-value component being a dominant contributor to value, but not with the private-value component being a dominant contributor to bidder value.

\subsection{Arbitrage Value and Price Volatility}
\begin{figure}[t]
        \centering
        \includegraphics[width=0.99\linewidth]{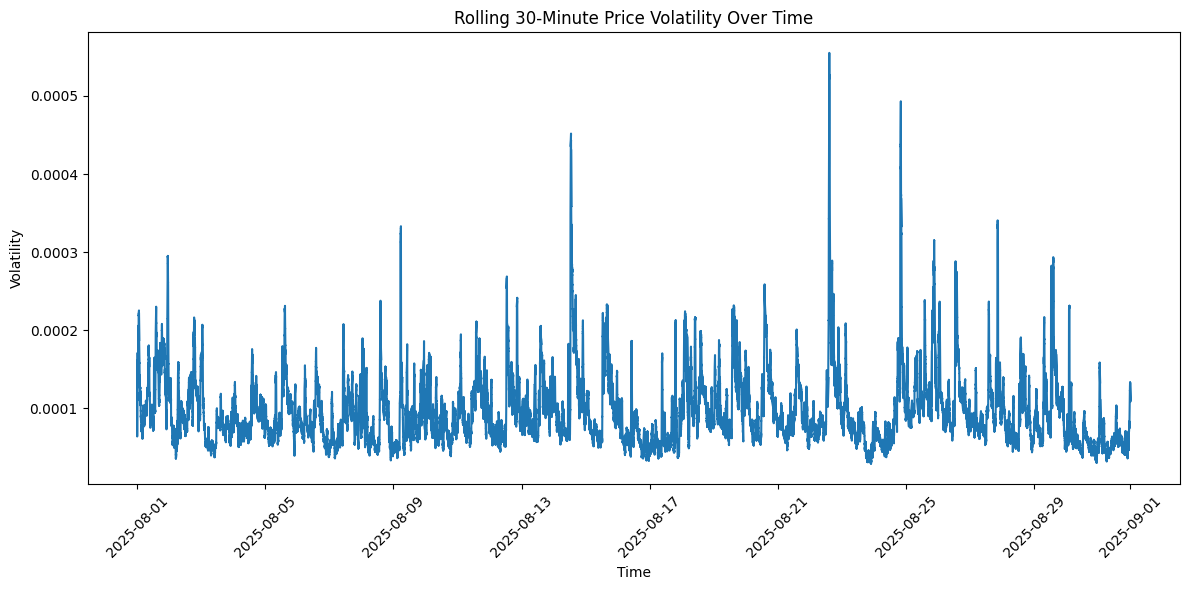}
        \caption{Price Volatility}
        \label{fig:arbitrage}
\end{figure}

The theory of arbitrage in Automated Market Makers~\cite{milionis2022automated} can inform how much arbitrage value searchers can expect to extract through a fast lane interval. While the theory applies to the case where bidders do not face uncertainty about the amount of arbitrage their individual trades will extract~\footnote{Under the TimeBoost policy, the fast lane transactions do not know the exact state of the AMMs as unobserved slow lane transaction might change it compared to the state that the arbitrageur last observed.}, quantitatively we should expect similar results. The theory predicts that arbitrage value is proportional to price variance, i.e. the arbitrage gains over a period of length $T$ should be proportional to the price variance over a period of length $T$,
$$ARB_T\sim\sigma^2_T.$$
Volatility in cryptocurrency price data is typically not stationary and is best modeled
by a mean-reverting stochastic volatility process. See Figure~\ref{fig:arbitrage}.
The theory makes the following predictions:
\begin{itemize}
\item If the price variance is linear in time, then arbitrage gains are linear in the length of the TimeBoost interval.
\item Arbitrage gains are mean-reverting. Longer run arbitage gain averages are easier to predict than short run arbitrage gains.
\end{itemize}

\section{Methodology}
\paragraph{Exchanges and Data}
We collected all TimeBoosted transactions that trade on decentralized exchanges on Arbitrum chain, in August, 2025, which is the last full month at the time of writing this note. We matched the data with one second tick Binance price data\footnote{https://data.binance.vision/} for the same time period. We moreover, collected bidding data for the TimeBoost auction for the same period.

The transaction data for this study is obtained from the Arbitrum blockchain transaction records accessible through Dune Analytics for the month of August. The auction data was restricted to non-empty winning rounds involving the three main participants, Selini Capital, Wintermute, and Kairos. There were other bidders that occasionally participated in the auction, but their combined share has been less than $0.1\%$. The restriction was applied to enable the linkage between auctions and transaction-level data. For each selected auction, information was collected on the round number, start and end times, winning participant, highest bid, and final paid bid in ETH, resulting in a data set of 43,248 auction rounds.
For the transaction-level analysis, only Arbitrum TimeBoosted transactions containing decentralized exchange (DEX) swaps were considered, reflecting the focus on CEX-DEX arbitrage activity. It is easy to check that these swaps are not cyclic arbitrage and there is only a single swap happening on chain. The data set was further refined to include decentralized exchanges on Arbitrum chain with the most liquid pairs, which are frequently traded by parties with the TimeBoost fast lane: WETH/USDC, WETH/USDT, WBTC/USDC, WBTC/USDT, WBTC/WETH.

This filtering yields 3,006,396 transactions with an estimated trading volume of 6.65 billion USD, over the study period. The total number of TimeBoost transactions in August is about 4.1M, while the total volume is 7.9 billion USD. Hence, CEX-DEX arbitrage accounts for overwhelming majority of TimeBoost transactions, both in number and volume.
The swap events include activity across multiple DEXs, including Uniswap (v2, v3, and v4), PancakeSwap (v2 and v3), Fluid (v1), Camelot (v2 and v3), SushiSwap (v1 and v2), as well as a range of smaller DEXs. Curve is the only major DEX excluded from the analysis due to the absence of relevant swap events in the dataset.

\paragraph{Markouts and Profits}
For each exchange transaction $t$ in our data set that exchanges tokens $A$ into tokens $B$, we keep track of how many tokens $t_A$ were exchanged for how many tokens $t_B$, net of pool fees. This gives an average price $\frac{t_B}{t_A}$ for token $t_B$ in token $t_A$. Each transaction $t$ also has a timestamp $t_T$ and the auction winner who submitted it, denoted by $t_W$. We approximate the arbitrage profit of a DEX trade, by the revenue that the
searcher can make by flattening their DEX-acquired inventory on the CEX,
net of any trading fees. As the searcher’s actual attainable execution price on
CEX is not observable, we estimate their revenue by using markouts calculated
with CEX mid prices at different mark out times.  That is, we check the  mid-price of tokens $A$ and $B$ on Binance (measured in $\$$) at time $t_T+m_{t_W}$ for a markout time $m_{t_W}$, denoted by $p_{A}(t_T+m_{t_W})$ resp. $p_{B}(t_T+m_{t_W})$, and obtain potential profit $$\Pi(t):=p_{A}(t_T+m_{t_W})t_A-p_B(t_T+m_{t_W})t_B-t_{fee},$$ where $t_{fee}$ is the chain gas fees the transaction pays. We calculate profit for different markout times and report the optimal (with second granularity) one which is 5 seconds. Empirical findings are not significantly affected by changing markout time in the interval of $[2s, 10s]$.

We distinguish between two types of profit statistics: in the first type, we aggregate all transaction markouts, in the second, we aggregate markouts by excluding potential negative profits.
This is motivated by the fact that participants likely have idiosyncratic reasons for fast lane usage and therefore some transactions are not for the purpose of arbitrage extraction. For example, sometimes they trade because they are re-balancing their portfolios. This is done as part of their overall risk management strategy, as some of the fast lane users are market makers trading on different platforms. 
Overall, we find that estimated performance improves significantly by excluding losses, which is a further motivation for such filtration.  Out of total 3.006M transactions, around 2.3M transactions result in positive markouts.

\section{Empirical Results}
Overall, the estimated arbitrage profits from the fast lane transaction are highly variable with few high volatility days (and minutes) contributing disproportionally to overall profits. See Figure~\ref{fig:placeholder}.

Among the three auction participants, we observe different usage of the fast lane. Wintermute and Selini actively use the fast lane with Wintermute exhibiting higher transaction frequency:
\begin{center}
\begin{tabular}{ cccccc } 
 \hline
 Winner & Total Txs & Total Rounds & Avg. Txs & Total Gas & Tot. Gas Fees\\ 
 \hline 
 Wintermute & 2.262M & 21237 & 106.5 & 3.88e+11 & 43563 \\ 
 Selini & 855K & 20831 & 41 & 1.32e+11 & 15803\\
 Kairos & 4.1K & 1180 & 3.5 & 5.72e+8 & 47 \\
 \hline
\end{tabular}
\end{center}
Note that gas fees are denominated in USD, as well as profits/markouts in the following. After subtracting paid bids from the minute markouts, the profits are positive and significant for Wintermute and Selini in absolute terms as well as on average per auction.   Wintermute is able to extract more profit from the system. Kairos is an outlier, by hardly trading actively and making a loss on chain.  However, as Kairos is a product that resells the fast lane access to third parties, this is likely an underestimation of their profit from the system.

\begin{center}
\begin{tabular}{ ccccc } 
 \hline
 Winner & Unfilt. Markouts & Filt. Markouts & Total Paid & Total Bid \\ 
 \hline 
 Wintermute & 578K & 1.275M & 342K & 656K \\ 
 Selini & 496K &  865K & 237K & 376K \\
 Kairos & 1.1K & 1.5K & 20K & 39K \\
 \hline
\end{tabular}
\end{center}

\subsection{Profits, Price Volatility and Bids}
Bids are a noisy predictor of arbitrage profits. More precisely, the correlation coefficients between profits and the winning (highest) bid and paid (second highest) bid is significant but low.
The following is a correlation table with filtered transactions: 

\begin{center}
\begin{tabular}{ c | cc | cc }
 \hline
 Winner & \multicolumn{2}{c|}{Pearson} & \multicolumn{2}{c}{Spearman} \\ 

        & Highest & Paid & Highest & Paid \\
 \hline
 Wintermute & 0.32 & 0.33 & 0.37 & 0.38 \\ 
 Selini     & 0.26 & 0.25 & 0.35 & 0.33 \\
 Kairos     & 0.17 & 0.18 & 0.21 & 0.21 \\
 \hline
\end{tabular}
\end{center}

The correlation is even lower for the unfiltered data set. This justifies using filtered markouts for a profit estimation, instead of unfiltered (all) markouts.

\begin{center}
\begin{tabular}{ c | cc | cc }
 \hline
 Winner & \multicolumn{2}{c|}{Pearson} & \multicolumn{2}{c}{Spearman} \\ 

        & Highest & Paid & Highest & Paid \\
 \hline
 Wintermute & 0.19 & 0.17 & 0.05 & 0.05 \\ 
 Selini & 0.16 & 0.16 & 0.06 & 0.06\\
 Kairos & 0.15 & 0.15 & 0.05 & 0.05\\
 \hline
\end{tabular}
\end{center}

\begin{figure}[t]
    \centering
    \includegraphics[width=0.99\linewidth]{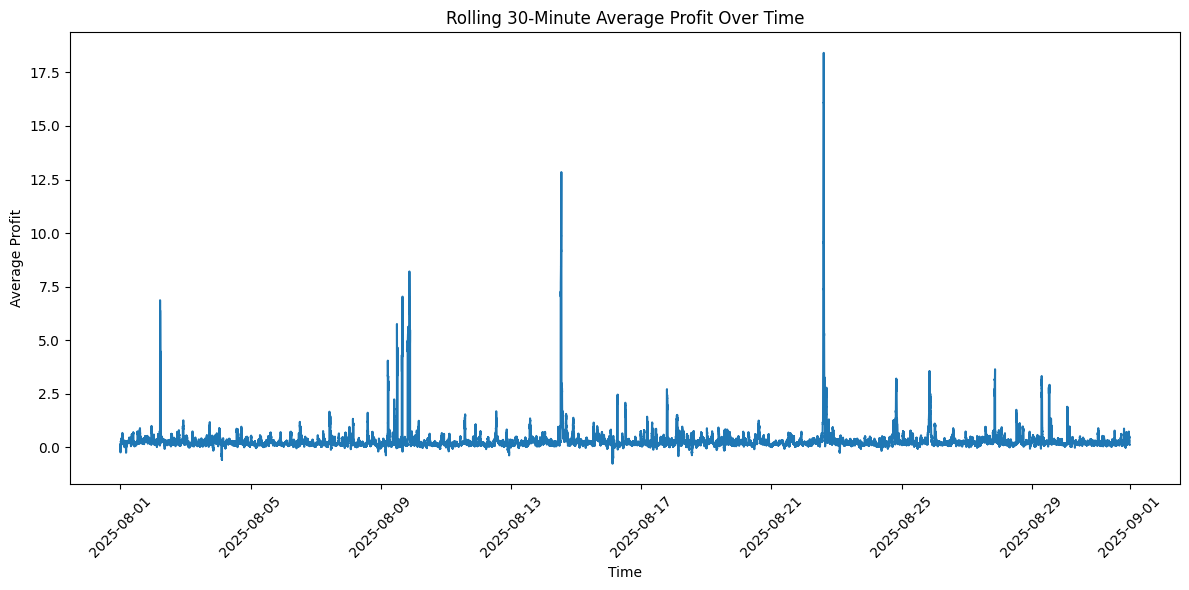}
    \caption{Arbitrage Profits}
    \label{fig:placeholder}
\end{figure}

One theory to make sense of the predictive power of bids was that arbitrage profits are best prediced by return variance.  
Return variance of ETH-USDC is a modestly good predictor of estimated arbitrage profits for the pair, see the regression table~\ref{table}. The fit however, is much better over longer time horizons (for which the data has higher $R^2$) than the minute intervals of the TimeBoost auction. The top bid (similarly the paid bid) are very noisy predictors of mark-out profit. 
\begin{table}[t]
\centering
\begin{tabular}{lccc}
\hline\hline
 & (1) & (2) & (3) \\
\hline
Constant & -0.565 & 9.999*** & 0.137 \\
         & (0.826) & (1.137) & (1.114) \\

Variance per minute & 8.773*** & & \\
         & (0.072) & & \\

Paid bid & & 1.395*** & \\
               & & (0.046) & \\

Paid bid next round & & & 2.120*** \\
        & & & (0.045) \\
\hline
Observations & 30,036 & 30,036 & 30,036 \\
$R^2$        & 0.333  & 0.030 & 0.070 \\
Adj.\ $R^2$  & 0.333  & 0.030 & 0.070 \\
F-statistic  & 15020.0 & 936.8 & 2253.0 \\
\hline\hline
\multicolumn{4}{l}{\footnotesize Standard errors in parentheses.} \\
\end{tabular}
\caption{OLS Regression Results for WETH-USDC pools}\label{table}
\end{table}

This suggests to look at the correlations of bids and arbitrage profits on other time interval lengths.
Next, we look at shorter time intervals and correlations between markouts in a shorter time window and winning bids resp. paid amounts. In particular, we take the first half or first quarter of the minute for which the bid is made. We focus on filtered transactions and the corresponding markouts.
The (Pearson) correlation table for quarter- half- minute markouts and different types of bids:

\begin{center}
\begin{tabular}{ c | cc | cc }
 \hline
 Winner & \multicolumn{2}{c|}{15 seconds} & \multicolumn{2}{c}{30 seconds} \\ 

        & Highest & Paid & Highest & Paid \\
 \hline 
 Wintermute & 0.24 & 0.23  & 0.29 & 0.28  \\
 Selini & 0.15 & 0.15 &  0.23 & 0.23 \\
 Kairos & 0.12 & 0.1 & 0.13 & 0.14 \\
 \hline
\end{tabular}
\end{center}

Interestingly, the correlation with bids decreases for the shorter time intervals, even though they end earlier relative to the auction end. One potential explanation of this counterintuitive finding is so called ''cold start" problem: in the first 5 second interval, the average share of fast lane transactions is $7.1\%$, while the share of aggregate markouts is $5.8\%$ for unfiltered and $6.2\%$ for filtered. Assuming the value and the number of opportunities is uniformly distributed, these shares must approximately be equal to $8.3\%$.  

On the other hand, bids are a good predictor of longer run arbitrage profits, which is consistent with our theoretical predictions of lower volatility on longer time intervals.

\begin{center}
\begin{tabular}{ c|ccccc } 
 \hline
 Bid type & 5 minutes & 10 minutes & 15 minutes & 30 minutes & 1h\\ 
 \hline 
 Highest & 0.62 & 0.77 & 0.78 & 0.83 & 0.877 \\ 
 Paid & 0.6 &  0.78 & 0.8 & 0.82 & 0.884 \\
 \hline
\end{tabular}
\end{center}

Another explanation of high correlation on longer interval is a simple bidding strategy players seem to exhibit. The strategy is to base the next bid on the previous minute profits the player (or its competitor)  makes through fast lane usage. To corroborate this behavior, we correlate bids with previous minute profits, current minute profits and future minute profits (aggregated over players). All numbers are for filtered transactions. The correlation is highest with previous profits, and worst with future profits:

\begin{center}
\begin{tabular}{ c| cc | cc }
 \hline
 bid &  \multicolumn{2}{c|}{Pearson} & \multicolumn{2}{c}{Spearman} \\ 

    &   filtered & unfiltered & filtered & unfiltered \\
 \hline
  next period & 0.50  & 0.4 & 0.52 & 0.43 \\
 current period&0.27  & 0.25& 0.35 & 0.35 \\
 previous period & 0.22& 0.20 & 0.31  & 0.35 \\
 \hline
\end{tabular}
\end{center}

\begin{figure}[t]
    \centering
    \includegraphics[width=0.45\linewidth]{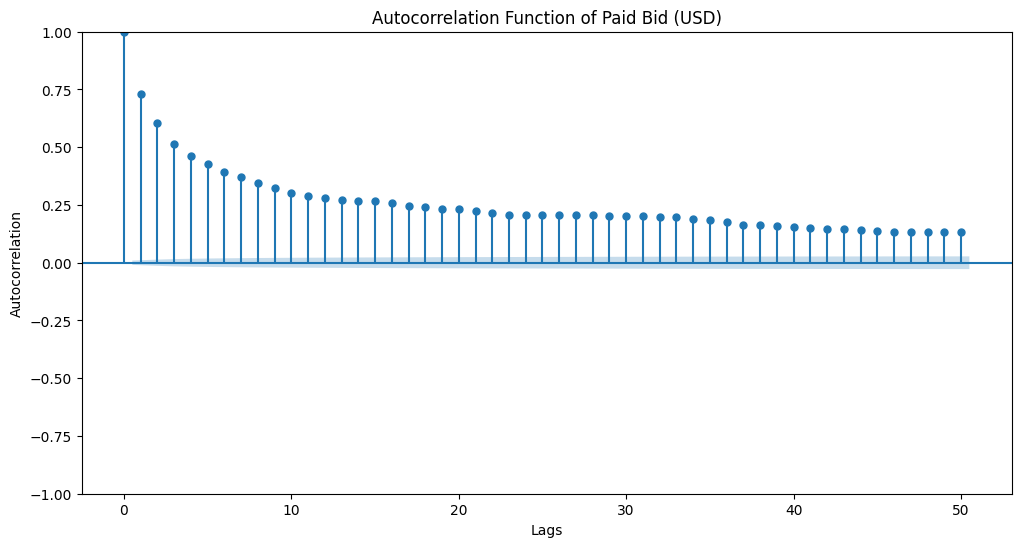}
    \includegraphics[width=0.45\linewidth]{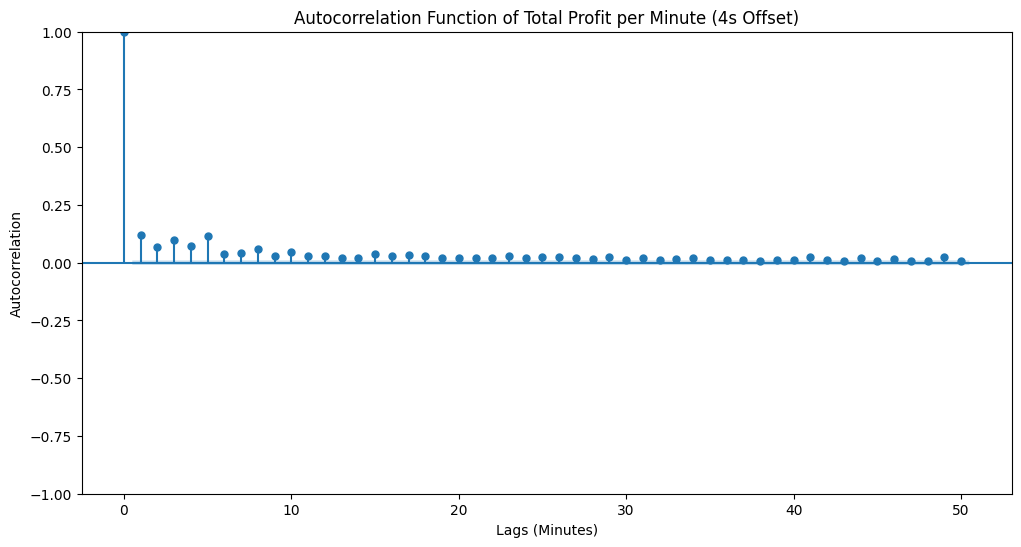}
    \caption{Autocorrelation of paid bid and of markouts per minute}
    \label{fig:placeholder}
\end{figure}

\begin{comment}
Autocorrelation table between positive markouts, all markouts, highest bids and paid bids: 
\begin{center}
    \begin{tabular}{ccccc}
    \hline 
    Shift & Positive Markouts & All Markouts &  Highest bid & Paid bid \\
    1 & 0.2900771 & 0.17 & 0.6618091 & 0.72933 \\
    2 & 0.1195192 & 0.002& 0.5091812 & 0.604794 \\
    3 & 0.1326923 & 0.0467 & 0.439213 & 0.515035 \\
    4 & 0.1330094 & 0.068 & 0.4013974 & 0.462621 \\
    5 & 0.1158865 & 0.0547 & 0.3927115 & 0.428626 \\
    6 & 0.07036826 & 0.021 & 0.3872046 & 0.393886 \\
    \hline
    \end{tabular}
\end{center}

7 markout 0.06886887 winning bids 0.3624387 paid bids 0.372449
8 markout 0.07486958 winning bids 0.3384778 paid bids 0.346476
9 markout 0.09946929 winning bids 0.3228069 paid bids 0.323405    
\end{comment}
The autocorrelation of the highest or paid bids is quite significant, see Figure~3, which is expected as there are long trends in bidding behavior, corresponding to market sentiment. On the other hand, autocorrelation between markouts is quite low. The only significant correlation is between consecutive positive markout values.

\begin{center}
    \begin{tabular}{cc}
    \hline 
    Shift\quad & 30s Filtered    \\    \hline 
    1 & 0.260  \\
    2 & 0.183  \\
    3 & 0.099  \\
    4 & 0.069 \\
    5 & 0.071  \\
    6 & 0.090 \\
    \hline
    \end{tabular}
\end{center}

\section{Conclusions and Future Work}
Analyzing TimeBoost auction data shows that estimating immediate future arbitrage value is hard.  Bids are a reasonably good predictor of longer time average profits, but a very noisy predictor of one minute profits.  

Insights obtained in this note can be used to make some predictions about the likely performance of slot auctions on the Ethereum chain, as they can also be considered as AOT auctions. Given that predicting arbitrage profits seems to be a difficult task even for short time intervals, auction revenue and efficiency is likely to suffer from switching from block to slot auctions. This casts some doubts on the idea of AOT slot auctions as a market structure for block building.
This work only scratches the surface of TimeBoost fast lane usage and auction. It is interesting to check how TimeBoost affects liquidity providers and the performance of other searchers that do not compete in the TimeBoost auction.

\bibliographystyle{plain}
\bibliography{sample}
\appendix
\subsection*{Data on an Extended Time Interval}

In this section, we will document the data that span a three-month period, June to August. Unlike the preceding section, data in this section is obtained from an Arbitrum full node, executing all transactions. Due to the stability issue with the Arbitrum node we run, some blocks are not indexed. In total, of 31,776,014 blocks from 342,607,969 to 374,383,982, we have indexed 28,807,014, which accounts for 90.65\% of all blocks.
The following table shows the total number of transactions, number of rounds, and average number of transactions per auction winner per round:
\begin{center}
    \begin{tabular}{cccc}
        \hline
        Player & Total number & Total rounds & \# of Txs per round \\
        \hline
        Wintermute & 4.69M & 56285 & 83.32 \\
        Selini & 2.20M & 66686 & 33.03 \\
        Kairos & 8.89K & 7636 & 1.16 \\
        \hline
    \end{tabular}
\end{center}

Table with total markouts and bids/paid amounts across players:
\begin{center}
    \begin{tabular}{ccccc}
        \hline
        Player & Unfiltered markouts & Filtered markouts & Total paid & Total bid \\
        \hline
        Wintermute & 863K & 2.152M & 1.138M & 1.848M \\
        Selini  & 1.074M & 2.070M & 1.087M & 1.718M \\
        Kairos & 4.50K & 5.745K & 160K & 281K \\
        \hline
    \end{tabular}
\end{center}
Relative ordering of players across their activity, volumes and total bids is preserved over the long time interval as well.

Correlation table with filtered transactions:
\begin{center}
\begin{tabular}{ ccccc } 
 \hline
 Player & Pearson Highest & Pearson Paid & Spearman Highest & Spearman Paid \\ 
 \hline 
 Wintermute & 0.2311 & 0.2015 & 0.3242 & 0.2994 \\ 
 Selini & 0.2375 & 0.2239 & 0.3526 & 0.3409 \\
 Kairos & 0.0896 & 0.0912 & 0.1862 & 0.1941 \\
 \hline
\end{tabular}
\end{center}

On a longer time interval Selini has a slightly better performance than Wintermute, which can be explained by a fact that the former is an earlier player in the TimeBoost auction.

Correlation table with all transactions:
\begin{center}
\begin{tabular}{ ccccc } 
 \hline
 Player & Pearson Highest & Pearson Paid & Spearman Highest & Spearman Paid \\ 
 \hline 
 Wintermute & 0.1321 & 0.1066 & 0.1254 & 0.1122 \\ 
 Selini & 0.1528 & 0.1448 & 0.1377 & 0.1474 \\
 Kairos & 0.0741 & 0.0726 & 0.0501 & 0.0535 \\
 \hline
\end{tabular}
\end{center}

The data is similar to August, correlations between unfiltered markouts and bids is considerably lower than between filtered markouts and bids.

Correlation table for half-minute and filtered transactions:
\begin{center}
\begin{tabular}{ ccccc } 
 \hline
 Player & Pearson Highest & Pearson Paid & Spearman Highest & Spearman Paid \\ 
 \hline 
 Wintermute & 0.2292 & 0.1895 & 0.3122 & 0.2891 \\ 
 Selini & 0.1859 & 0.1747 & 0.3424 & 0.3299 \\
 Kairos & 0.0770 & 0.0736 & 0.1574 & 0.1644 \\
 \hline
\end{tabular}
\end{center}

Correlation table for quarter-minute and filtered transactions:
\begin{center}
\begin{tabular}{ ccccc } 
\hline
 Player & Pearson Highest & Pearson Paid & Spearman Highest & Spearman Paid \\ 
 \hline 
 Wintermute & 0.1580 & 0.1267 & 0.2975 & 0.2794 \\ 
 Selini & 0.1260 & 0.1157 & 0.3113 & 0.3001 \\
 Kairos & 0.0686 & 0.0688 & 0.1220 & 0.1315 \\
 \hline
\end{tabular}
\end{center}

The correlation coefficients are similar to those of August. 
The effect of dropping correlation on shorter time intervals is present on a longer time horizon as well.

\end{document}